\documentclass[12pt,reqno,a4paper]{article}
\usepackage{amsmath,amsfonts,amssymb,graphicx,epsfig,pict2e,rotating}
\usepackage[english]{babel}
\usepackage[noxcolor]{pstricks}
\usepackage{hyperref}
\hypersetup{backref,
colorlinks=true,
citecolor=darkblue}
\numberwithin{equation}{section}
\definecolor{darkblue}{rgb}{0,0,.8}
\definecolor{lightblue}{rgb}{.65,.95,1}
\definecolor{lightlightblue}{rgb}{.85,1,1}
\definecolor{rred}{rgb}{1,0,0}

\newcommand{\be}{\begin{equation}}
\newcommand{\ee}{\end{equation}}
\newcommand{\bea}{\begin{eqnarray}}
\newcommand{\eea}{\end{eqnarray}}
\newcommand{\D}{{\mathbf D}}

\newcommand{\uno}{{\mathbf 1}}

\def\vec#1{\boldsymbol{#1}}
\newcommand{\spos}[2]{\makebox(0,0)[#1]{$\small{#2}$}}
\newcommand{\sposb}[2]{\makebox(0,0)[#1]{$ #2 $}}

\newcommand{\ba}[1]{\begin{array}{@{}#1@{}}}
\newcommand{\ea}{\end{array}}
\newcommand{\bpic}{\begin{picture}}
\newcommand{\epic}{\end{picture}}

\def\hbar{{\overline{h}}}

\def\binom#1#2{\left(#1\atop #2\right)}

\def\({\left(}
\def\){\right)}
\font\tenmsb=msbm10 scaled \magstep1
\font\sevenmsb=msbm7 scaled \magstep1
\font\fivemsb=msbm5 scaled \magstep1
\newfam\msbfam
\textfont\msbfam=\tenmsb
\scriptfont\msbfam=\sevenmsb
\scriptscriptfont\msbfam=\fivemsb

\long\def\omit#1{}

\def\nar#1#2#3{\mbox{ $\left<#1\atop #2\thinspace ,\thinspace #3\thinspace\right>_q$}}

\def\smaller{\small}

\def\emptysquare{\hspace{-.11\unitlength}
\begin{pspicture}(1,1)
\pspolygon[linewidth=.25pt](0,0)(1,0)(1,1)(0,1)(0,0)
\end{pspicture}}
\def\loopa{\hspace{-.11\unitlength}
\begin{pspicture}(1,1)
\pspolygon[linewidth=.25pt](0,0)(1,0)(1,1)(0,1)(0,0)
\psarc[linewidth=1.5pt](1,0){.5}{90}{180}
\psarc[linewidth=1.5pt](0,1){.5}{-90}{0}
\end{pspicture}}
\def\loopb{\hspace{-.11\unitlength}
\begin{pspicture}(1,1)
\pspolygon[linewidth=.25pt](0,0)(1,0)(1,1)(0,1)(0,0)
\psarc[linewidth=1.5pt](0,0){.5}{0}{90}
\psarc[linewidth=1.5pt](1,1){.5}{180}{270}
\end{pspicture}}

\def\down2monoid#1#2#3#4{\rule[-4\unitlength]{0in}{8\unitlength}
\begin{picture}(0,0)(-#1,-#2)
\put(2,0){\oval(12,10)[t]}
\put(-4,-1){\makebox(0,0)[t]{\smaller \mbox{$#3$}}}
\put(8,-1){\makebox(0,0)[t]{\smaller \mbox{$#4$}}}
\end{picture}}

\def\pdiamonda{
\begin{pspicture}(0,0)(0,8)
\pspolygon[linewidth=.25pt](0,4)(2,0)(4,4)(2,8)
\psbezier[linewidth=1pt](1,2)(2.1,2.5)(2.1,5.5)(1,6)
\psbezier[linewidth=1pt](3,2)(1.9,2.5)(1.9,5.5)(3,6)
\end{pspicture}}
\def\pdiamondb{
\begin{pspicture}(0,0)(0,8)
\pspolygon[linewidth=.25pt](0,4)(2,0)(4,4)(2,8)
\psarc[linewidth=1pt](2,0){2.236}{63.4}{116.5}
\psarc[linewidth=1pt](2,8){2.236}{-116.5}{-63.4}
\end{pspicture}}

\def\psq#1{
\begin{pspicture}(0,1)(4,5)
\pspolygon[linewidth=.25pt](0,0)(4,0)(4,4)(0,4)
\rput(2,2){\small $#1$}
\psarc(0,0){.35}{0}{90}
\end{pspicture}}

\def\psqa#1{
\begin{pspicture}(0,1)(4,5)
\pspolygon[linewidth=.25pt](0,0)(4,0)(4,4)(0,4)
\psarc[linewidth=1pt](4,0){2}{90}{180}
\psarc[linewidth=1pt](0,4){2}{-90}{0}
\rput(2,2){\small $#1$}
\end{pspicture}}

\def\psqb#1{
\begin{pspicture}(0,1)(4,5)
\pspolygon[linewidth=.25pt](0,0)(4,0)(4,4)(0,4)
\psarc[linewidth=1pt](0,0){2}{0}{90}
\psarc[linewidth=1pt](4,4){2}{180}{270}
\rput(2,2){\small $#1$}
\end{pspicture}}

\makeatletter
\renewcommand{\@makecaption}[2]{
   \vskip\abovecaptionskip
   \sbox\@tempboxa{#1. #2}%
   \ifdim \wd\@tempboxa >\hsize
     #1. #2\par
   \else
     \global \@minipagefalse
     \hb@xt@\hsize{\hfil\box\@tempboxa\hfil}%
   \fi
   \vskip\belowcaptionskip}
\makeatother

\numberwithin{equation}{section}

\begin{document}
\setcounter{page}{1}

\vspace{8mm}
\begin{center}
{\Large {\bf The Baxter $Q$ Operator of Critical Dense Polymers}}

\vspace{10mm}
 {\Large Alessandro Nigro\footnote{Email: Alessandro.Nigro@mi.infn.it}}\\
 [.3cm]
  {\em Dipartimento di Fisica and INFN- Sezione di Milano\\Universit\`a degli Studi di Milano IVia Celoria 16, I-20133 Milano, Italy}\\[.4cm]

  \end{center}

\vspace{8mm}
\centerline{{\bf{Abstract}}}
\vskip.4cm
\noindent
We consider critical dense polymers ${\cal L}_{1,2}$, corresponding to a logarithmic conformal field theory with central charge $c=-2$.  An elegant decomposition of the Baxter $Q$ operator is obtained in terms of a finite number of lattice integrals of motion. All local, non local and dual non local involutive charges are introduced directly on the lattice and their continuum limit is found to agree with the expressions predicted by conformal field theory. A highly non trivial operator $\Psi(\nu)$ is introduced on the lattice taking values in the Temperley Lieb Algebra. This $\Psi$ function provides a lattice discretization of the analogous function introduced by Bazhanov, Lukyanov and Zamolodchikov. It is also observed how the eigenvalues of the $Q$ operator reproduce the well known spectral determinant for the harmonic oscillator in the continuum scaling limit.

\renewcommand{\thefootnote}{\arabic{footnote}}
\setcounter{footnote}{0}

\section{Introduction}
It is well estabilished that a lattice approach to logarithmic minimal models \cite{PRZ}\cite{salLCFT} can be realized in terms of indecomposable representations of the Temperley Lieb Algebra \cite{Jones}, in particular the integrability of these lattice realizations of logarithmic CFTs is proved by the existence of commuting families  of double row $N-$tangles, the parameter of such a family being called the spectral parameter.\\
In contrast with unitary minimal models, which are realized on the lattice for example by the RSOS models \cite{bp}\cite{kp}\cite{nigro}, the transfer matrix may exhibit a Jordan indecomposable structure for some choice of Cardy-type boundary conditions. \\
It is well known \cite{solvpol}\cite{salll} that the CFT corresponding to critical dense polymers has central charge $c=-2$. Such a conformal field theory is known to be logarithmic, these theories, in contrast with rational CFTs, can be realized by different models for the same value of the central charge and conformal weights. For example Hamiltonian walks on a Manhattan lattice \cite{DuplantierDavid,Sedrakyan}, the rational triplet theory\cite{GabK,FeiginEtAl,GabRunk}, symplectic fermions \cite{Kau95,Kau00}, the Abelian sanpile model \cite{Ruelle}, dimers \cite{dimers}, the traveling salesman problem \cite{JRS} and branching polymers \cite{branchpoly} all share the same value of the central charge, which is $-2$. \\
In this paper we want to continue our analysys \cite{nigropol} of the Logarithmic Minimal Model ${\cal L}_{1,2}$, also called Critical Dense Polymers \cite{solvpol} in view of the well known works of Bazhanov, Lukyanov and Zamolodchikov \cite{blz}. Our aim is to explicitly characterize the ${\bf Q}$ operator of Critical Dense Polymers explicitly in terms of tangles taking value in the Temperley Lieb Algebra. In particular it is well known that knowledge of the ${\bf Q}$ operator is useful in deriving Bethe Ansatz equations, and furthermore its asymptotic expansions are well known to be related to all the local and non local conservation laws of the model. It turns out in this case that the Bethe Ansatz equations can be trivially solved and that we can define all the local and non local conservation laws directly on the lattice, and relate them explicitly to commuting families of tangles defined in the Temperley Lieb Algebra. The eigenvalues of these lattice involutive charges behave asymptotically in $1/N$, $N$ being the system size, as the corresponding eigenvalues of the quantum operators of \cite{blz} defined in Conformal Field Theory.\\
The technique of the integral transform ${\bf \Psi}(\nu)$ of the  operator ${\bf Q}(x)$, is used directly on the lattice to introduce non trivial operators which can be treated formally as an analytic prolongation to arbitrary complex values of the spin $i\nu=2n-1$ of the local involutive charges, the eigenvalues of these non trivial operators are also computed exactly on the lattice and related to their continuum analogues. \\
It turns out also that by taking a suitable continuum limit we recover something that is well knonw in the continuum limit, namely the fact that the eigenvalues of the ${\bf Q}$ operator can be expressed in terms of spectral determinants of the harmonic oscillator.

\section{Critical Dense Polymers}
\psset{unit=.1in}
\setlength{\unitlength}{.1in}
We will consider in this paper an exactly solvable model of critical dense polymers on a square lattice \cite{solvpol}. The degrees of freedom are localized on elementary faces, which can be found in one of the following two configurations:
\be \psqa{} \ {\rm or} \ \psqb{}  \ee
where the arcs represent segments of the polymer. The elementary faces belong to the planar Temperley-Lieb algebra \cite{Jones}, and therefore satitfy the following simple equations:
\psset{unit=.075in}
\setlength{\unitlength}{.075in}
\bea
\begin{pspicture}(0,3)(8,11)
\psline[linestyle=dashed,dash=.5 .5,linewidth=.25pt](2.4,0)(6.4,0)
\psline[linestyle=dashed,dash=.5 .5,linewidth=.25pt](2.4,8)(6.4,8)
\rput(0,4){\pdiamondb}
\rput(4,4){\pdiamondb}
\psarc[linewidth=1pt](4.4,4){2.236}{63.4}{116.5}
\psarc[linewidth=1pt](4.4,4){2.236}{-116.5}{-63.4}
\end{pspicture}
\;=\;
\begin{pspicture}(0,3)(4,11)
\rput(0,4){\pdiamondb}
\end{pspicture}\ ,\qquad\qquad
\begin{pspicture}(0,3)(8,11)
\psline[linestyle=dashed,dash=.5 .5,linewidth=.25pt](2.4,0)(6.4,0)
\psline[linestyle=dashed,dash=.5 .5,linewidth=.25pt](2.4,8)(6.4,8)
\rput(0,4){\pdiamonda}
\rput(4,4){\pdiamonda}
\psarc[linewidth=1pt](4.4,4){2.236}{63.4}{116.5}
\psarc[linewidth=1pt](4.4,4){2.236}{-116.5}{-63.4}
\end{pspicture}
\;=\;\beta\;
\begin{pspicture}(0,3)(4,10)
\rput(0,4){\pdiamonda}
\end{pspicture}
\eea \\
where the dashed lines indicate that the corners and associated incident edges are identified.\\
The parameter $\beta$ represents the loop fugacity which, for critical dense polymers, is set to zero. This means that the polymer is not allowed to form closed loops. Therefore it passes twice through each face of the lattice, and in the continuum scaling limit it is dense or space filling, in the sense that its fractal dimension is $2$.\\
The transfer matrix is built out of local face operators or 2-tangles $X(u)$ and boundary 1-triangles.\\
The local face operators are defined diagrammatically in the planar TL algebra:
\be
X(u)=\psq{u}=\cos(u)\ \psqa{}+\sin(u)\ \psqb{}
\ee 
which means that the weights assigned to the elementary face conficgurations are
\be
W\Bigg( \psqa{} \Bigg)\;=\;\cos(u),\qquad\qquad W\Bigg(\ \psqb{} \Bigg)\;=\;\sin(u)
\ee
The local face operators satisfy the Yang-Baxter equation as well as an Inversion Identity.\\
The $(1,s)$ boundary 1-triangles are defined as  the following solutions to the boundary Yang Baxter equation \cite{PRZ}: 
\bea
\begin{pspicture}(0,4)(4,13.5)
\pspolygon[linewidth=.25pt](0,4)(4,0)(4,8)(0,4)
\rput(2.3,4){\small $\pm i\infty$}
\rput(2.3,9){\small $(1,s)$}
\end{pspicture}\ \ =\ \  
\begin{pspicture}(-8,4)(8,12)
\psline[linewidth=.25pt](-8,0)(4,0)
\psline[linewidth=.25pt](-8,4)(4,4)
\psline[linewidth=.25pt](-8,8)(4,8)
\psline[linewidth=.25pt](-4,0)(-4,8)
\psline[linewidth=.25pt](-8,0)(-8,8)
\psline[linewidth=.25pt](0,0)(0,8)
\psline[linewidth=.25pt](4,0)(4,8)
\psline[linewidth=.25pt](4,4)(8,8)
\psline[linewidth=.25pt](4,4)(8,0)
\psline[linewidth=.25pt](8,0)(8,8)
\psarc[linewidth=1pt](4,4){2}{-90}{90}
\psarc[linewidth=1pt](0,4){2}{-90}{90}
\psarc[linewidth=1pt](4,0){2}{90}{180}
\psarc[linewidth=1pt](4,8){2}{-180}{-90}
\psarc[linewidth=1pt](-4,4){2}{-90}{90}
\psarc[linewidth=1pt](0,0){2}{90}{180}
\psarc[linewidth=1pt](0,8){2}{-180}{-90}
\psarc[linewidth=1pt](-8,4){2}{-90}{90}
\psarc[linewidth=1pt](-4,0){2}{90}{180}
\psarc[linewidth=1pt](-4,8){2}{-180}{-90}
\psline[linestyle=dashed,dash=.5 .5,linewidth=.25pt](4,0)(8,0)
\psline[linestyle=dashed,dash=.5 .5,linewidth=.25pt](4,8)(8,8)
\rput(-6,0){\spos{}{\bullet}}
\rput(-2,0){\spos{}{\bullet}}
\rput(2,0){\spos{}{\bullet}}
\rput(-6,8){\spos{}{\bullet}}
\rput(-2,8){\spos{}{\bullet}}
\rput(2,8){\spos{}{\bullet}}
\put(-8,-.5){$\underbrace{\hspace{12\unitlength}}_{\mbox{$s-1$ columns}}$}
\end{pspicture}\label{generalsbdy}\\
\nonumber
\eea \\
The YBEs, supplemented by additional local relations, are sufficient to imply
commuting transfer matrices and integrability.  To work on a strip with fixed boundary conditions on the
right and left, we need to work with $N$ column Double-row Transfer Matrices represented schematically in the planar TL algebra by the $N$-tangle
\setlength{\unitlength}{13mm}
\psset{unit=13mm}
\bea
\vec D(u)\;
=\quad
\raisebox{-1.3\unitlength}[1.3\unitlength][
1.1\unitlength]{\begin{pspicture}(6.4,2.4)(0.4,0.1)
\multiput(0.5,0.5)(6,0){2}{\line(0,1){2}}
\multiput(1,0.5)(1,0){3}{\line(0,1){2}}
\multiput(5,0.5)(1,0){2}{\line(0,1){2}}
\multiput(1,0.5)(0,1){3}{\line(1,0){5}}
\put(1,1.5){\line(-1,2){0.5}}\put(1,1.5){\line(-1,-2){0.5}}
\put(6,1.5){\line(1,2){0.5}}\put(6,1.5){\line(1,-2){0.5}}
\multiput(1.5,1)(1,0){2}{\sposb{}{u}}\put(5.5,1){\sposb{}{u}}
\multiput(1.5,2)(1,0){2}{\sposb{}{\lambda\!-\!u}}
\put(5.5,2){\sposb{}{\lambda\!-\!u}}
\put(0.71,1.5){\sposb{}{\lambda\!-\!u\ \ \ }}\put(6.29,1.5){\sposb{}{u}}
\multiput(0.5,0.5)(0,2){2}{\makebox(0.5,0){\dotfill}}
\multiput(6,0.5)(0,2){2}{\makebox(0.5,0){\dotfill}}
\psarc(1,.5){.125}{0}{90}
\psarc(1,1.5){.125}{0}{90}
\psarc(2,.5){.125}{0}{90}
\psarc(2,1.5){.125}{0}{90}
\psarc(5,.5){.125}{0}{90}
\psarc(5,1.5){.125}{0}{90}
\psarc[linewidth=1pt](1,1.5){.5}{245}{270}
\psarc[linewidth=1pt](1,1.5){.5}{90}{115}
\psarc[linewidth=1pt](6,1.5){.5}{65}{90}
\psarc[linewidth=1pt](6,1.5){.5}{270}{295}
\end{pspicture}}\qquad
\label{DTM}
\eea
This schematic representation in the {\it planar} TL algebra needs to be interpreted
appropriately to write $\vec D(u)$ in terms of the generators of the {\it linear} TL 
algebra and to write down its associated matrix:
\psset{unit=.9cm}
\setlength{\unitlength}{.9cm}
\bea
\begin{pspicture}(-1,5)(6,16.5)

\psline[linewidth=.5pt](0,6)(1,5)
\psline[linewidth=.5pt](0,6)(4,10)
\psline[linewidth=.5pt](1,5)(5,9)
\psline[linewidth=.5pt](4,8)(3,9)
\psline[linewidth=.5pt](1,7)(2,6)
\psline[linewidth=.5pt](2,8)(3,7)
\psline[linewidth=.5pt](4,8)(3,9)
\psline[linewidth=.5pt](4,10)(5,9)
\psline[linewidth=.5pt](4,10)(0,14)
\psline[linewidth=.5pt](5,11)(1,15)
\psline[linewidth=.5pt](5,9)(5,11)
\psline[linewidth=.5pt](2,8)(3,7)
\psline[linewidth=.5pt](4,10)(5,11)
\psline[linewidth=.5pt](3,11)(4,12)
\psline[linewidth=.5pt](2,12)(3,13)
\psline[linewidth=.5pt](1,13)(2,14)
\psline[linewidth=.5pt](0,14)(1,15)
\rput(1,6){$u$}
\rput(2,7){$u$}
\rput(4,9){$u$}
\rput(4,11){$u$}
\rput(2,13){$u$}
\rput(1,14){$u$}
\rput(4.575,10){$u,\xi$}
\psarc(1,5){.15}{45}{135}
\psarc(2,6){.15}{45}{135}
\psarc(4,8){.15}{45}{135}
\psarc(4,10){.15}{45}{135}
\psarc(2,12){.15}{45}{135}
\psarc(1,13){.15}{45}{135}
\psline[linewidth=1pt](-.5,5.5)(-.5,14.5)
\psline[linewidth=1pt](.5,6.5)(.5,13.5)
\psline[linewidth=1pt](1.5,7.5)(1.5,12.5)
\psline[linewidth=1pt](2.5,8.5)(2.5,11.5)
\psline[linewidth=1pt](3.5,9.5)(3.5,10.5)
\psline[linewidth=1pt](1.5,4.5)(1.5,5.5)
\psline[linewidth=1pt](2.5,4.5)(2.5,6.5)
\psline[linewidth=1pt](3.5,4.5)(3.5,7.5)
\psline[linewidth=1pt](4.5,4.5)(4.5,8.5)
\psline[linewidth=1pt](1.5,14.5)(1.5,15.5)
\psline[linewidth=1pt](2.5,13.5)(2.5,15.5)
\psline[linewidth=1pt](3.5,12.5)(3.5,15.5)
\psline[linewidth=1pt](4.5,11.5)(4.5,15.5)
\psbezier[linewidth=1pt](-.5,4.5)(-.5,5)(.5,5)(.5,4.5)
\psbezier[linewidth=1pt](-.5,5.5)(-.5,5)(.5,5)(.5,5.5)
\psbezier[linewidth=1pt](-.5,14.5)(-.5,15)(.5,15)(.5,14.5)
\psbezier[linewidth=1pt](-.5,15.5)(-.5,15)(.5,15)(.5,15.5)
\rput(-1,4.25){$j=$}
\rput(0,4.25){$-1$}
\rput(1,4.25){$0$}
\rput(2,4.25){$1$}
\rput(3,4.25){$\ldots$}
\rput(4,4.25){$N\!-\!1$}
\rput(5,4.25){$N$}
\rput(-2.5,10){$\vec{D}(u)=$}
\end{pspicture}
\label{expandedD}
\eea \\
or algebraically:
\begin{equation}
\vec D(u)=
{\bf e}_{-1}\Big(\prod_{j=0}^{N-1} X_j(u)\Big) K_s(u,\xi) \Big(\prod_{j=N-1}^0 X_j(u)\Big){\bf e}_{-1}
\label{TLDTM}
\end{equation}
being
\be X_j(u)=\uno_j \cos(u)+{\bf e}_j \sin(u)  \ee
\psset{unit=.055in}
\setlength{\unitlength}{.075in}
\be \uno_i= \begin{pspicture}(0,3)(4,11)
\rput(0,4){\pdiamondb}
\end{pspicture}\ \ee

\be  {\bf e}_i=\begin{pspicture}(0,3)(4,10)
\rput(0,4){\pdiamonda}
\end{pspicture}   \ee
where ${\bf e}_{-1}$ is an auxiliary generator which proves useful to express the transfer matrix in the linear TemperleyLieb Algebra, the loop generated by such auxiliary generator when squaring the transfer matrix, has vanishing a value and one has to remove it by hand in this representation.\\ 
Clearly the ${\bf e}_j$ satisfy in the linear Temepley Lieb Algebra :
\be {\bf e}^2_j=0  \ee
\be  {\bf e}_{j\pm 1}  {\bf e}_j {\bf e}_{j\pm 1}= {\bf e}_{j\pm 1}  \ee
 
Moreover the boundary tangle $K_s(u,\xi)$ is proportional to the identity in the linear TL algebra for $(1,s)$ boundary conditions.\\
The matrix representation of the N-tangle is obtained by acting from below (or above) on a basis of link states with $s-1$ defects, for example the following represents a link state with 11 nodes and three defects
\psset{unit=.65cm}
\setlength{\unitlength}{.65cm}
\be
\begin{pspicture}(11,2)
\psarc[linewidth=1.5pt](1,0){.5}{0}{180}
\psline[linewidth=1.5pt](2.5,0)(2.5,1.5)
\psarc[linewidth=1.5pt](5,0){.5}{0}{180}
\psarc[linewidth=1.5pt](5,0){1.5}{0}{180}
\psarc[linewidth=1.5pt](8,0){.5}{0}{180}
\psline[linewidth=1.5pt](9.5,0)(9.5,1.5)
\psline[linewidth=1.5pt](10.5,0)(10.5,1.5)
\end{pspicture}
\label{link}
\ee

\begin{figure}[thbp]
\label{defectsTM}
\psset{unit=1.1cm}
\setlength{\unitlength}{1.1cm}
\begin{center}
\begin{pspicture}(10,7)
\pspolygon[fillstyle=solid,fillcolor=lightlightblue,linewidth=.25pt](0,0)(8,0)(8,4)(0,4)(0,0)
\pspolygon[fillstyle=solid,fillcolor=lightblue,linewidth=.25pt](8,0)(10,0)(10,4)(8,4)(8,0)
\rput(0,5){$\color{black}(r',s')=(1,1)$}
\rput(10.6,5){$\color{black}(r,s)=(1,3)$}
\rput[bl](8,0){\emptysquare}
\rput[bl](9,0){\emptysquare}
\rput[bl](8,1){\emptysquare}
\rput[bl](9,1){\emptysquare}
\rput[bl](8,2){\emptysquare}
\rput[bl](9,2){\emptysquare}
\rput[bl](8,3){\emptysquare}
\rput[bl](9,3){\emptysquare}
\psline[linecolor=blue,linestyle=dashed,dash=.25 .25,linewidth=2pt](0,4)(10,4)
\psline[linecolor=blue,linewidth=1.5pt](8.5,0)(8.5,4)
\psline[linecolor=blue,linewidth=1.5pt](9.5,0)(9.5,4)
\psline[linecolor=blue,linewidth=1.5pt](8,.5)(8.4,.5)
\psline[linecolor=blue,linewidth=1.5pt](8.6,.5)(9.4,.5)
\psline[linecolor=blue,linewidth=1.5pt](9.6,.5)(10,.5)
\psline[linecolor=blue,linewidth=1.5pt](8,1.5)(8.4,1.5)
\psline[linecolor=blue,linewidth=1.5pt](8.6,1.5)(9.4,1.5)
\psline[linecolor=blue,linewidth=1.5pt](9.6,1.5)(10,1.5)
\psline[linecolor=blue,linewidth=1.5pt](8,2.5)(8.4,2.5)
\psline[linecolor=blue,linewidth=1.5pt](8.6,2.5)(9.4,2.5)
\psline[linecolor=blue,linewidth=1.5pt](9.6,2.5)(10,2.5)
\psline[linecolor=blue,linewidth=1.5pt](8,3.5)(8.4,3.5)
\psline[linecolor=blue,linewidth=1.5pt](8.6,3.5)(9.4,3.5)
\psline[linecolor=blue,linewidth=1.5pt](9.6,3.5)(10,3.5)
\rput[bl](0,0){\loopb}
\rput[bl](1,0){\loopb}
\rput[bl](2,0){\loopa}
\rput[bl](3,0){\loopa}
\rput[bl](4,0){\loopa}
\rput[bl](5,0){\loopb}
\rput[bl](6,0){\loopa}
\rput[bl](7,0){\loopa}
\rput[bl](0,1){\loopa}
\rput[bl](1,1){\loopa}
\rput[bl](2,1){\loopa}
\rput[bl](3,1){\loopa}
\rput[bl](4,1){\loopa}
\rput[bl](5,1){\loopa}
\rput[bl](6,1){\loopb}
\rput[bl](7,1){\loopb}
\rput[bl](0,2){\loopb}
\rput[bl](1,2){\loopb}
\rput[bl](2,2){\loopb}
\rput[bl](3,2){\loopb}
\rput[bl](4,2){\loopb}
\rput[bl](5,2){\loopb}
\rput[bl](6,2){\loopb}
\rput[bl](7,2){\loopa}
\rput[bl](0,3){\loopa}
\rput[bl](1,3){\loopa}
\rput[bl](2,3){\loopa}
\rput[bl](3,3){\loopa}
\rput[bl](4,3){\loopa}
\rput[bl](5,3){\loopa}
\rput[bl](6,3){\loopa}
\rput[bl](7,3){\loopa}
\psarc[linecolor=blue,linewidth=1.5pt](0,1){.5}{90}{270}
\psarc[linecolor=blue,linewidth=1.5pt](0,3){.5}{90}{270}
\psarc[linecolor=blue,linewidth=1.5pt](10,1){.5}{-90}{90}
\psarc[linecolor=blue,linewidth=1.5pt](10,3){.5}{-90}{90}
\psarc[linecolor=red,linewidth=1.5pt](1,4){.5}{0}{180}
\psarc[linecolor=red,linewidth=1.5pt](7,4){.5}{0}{180}
\psarc[linecolor=red,linewidth=1.5pt](4,4){.5}{0}{180}
\psarc[linecolor=red,linewidth=1.5pt](7,4){1.5}{0}{180}
\psarc[linecolor=red,linewidth=1.5pt](6,3.105){3.62}{14}{166}
\end{pspicture}
\caption{A typical configuration on the strip showing connectivities. The action on the link state is explained in the next section. The boundary condition is of type $(r',s')=(1,1)$ on the left and type $(r,s)=(1,3)$ on the right so there are $\ell=s\!-\!1=2$ defects in the bulk. The strings propagating along the right boundary are spectators connected to the defects.}
\end{center}
\end{figure}
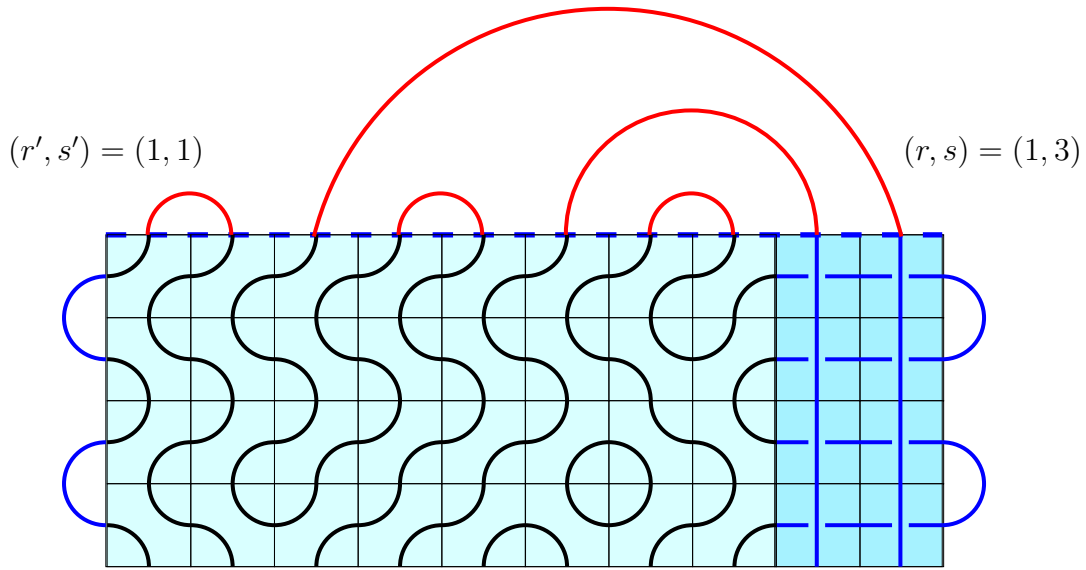
For $(1,s)$ boundary conditions the transfer matrix acts on link states with $\ell=s-1$ defects which have to be glued into the $(1,s)$ boundary triangle as exeplified in figure \ref{defectsTM}.

\subsection{Inversion Identities}
We first of all set some notation, let $x=\sin(2u)$, then we have that for $(1,s)$ boundary conditions the tranfer matrix satisfies an inversion identity \cite{PRZ}, which by virtue of commutativity is satisfied also by its eigenvalues:
\be \label{inv} D(x)D(-x)=\mathcal{F}^2_N(x^2)\ee
Such an identity does not depend on $s$ and it can be solved exactly for finite $N$, yielding a number of solutions which is larger than the size of the $\D$ matrix.\\
The idea behind the solution is the observation that ${\cal F}_N(u)$ is an entire function of $u$ which can be factorized exactly. The eigenvalues $D(u)$ are determined by sharing out the zeroes of ${\cal F}$ between the two factors on the righthand side of (\ref{inv}).\\
The function ${\cal F}^2$, due to being a square, has only double zeroes which we can define through:
\be {\cal F}^2 _N(x^2_k)=0   \ee
where
\be x_k=\frac{1}{\sin t_k}  \ee
being $t_j=\frac{j\pi}{N}$ for even $N$ whereas $t_j=\frac{(2j-1)\pi}{2N}$ for odd $N$.\\
With this notation one has for even $N=2D+2$:
\be {\cal F}_N(x^2)=\frac{D+1}{2^{2D}}\prod_{k=1}^{D}(x^2_k-x^2)    \ee
whereas for odd $N=2D+1$:
\be {\cal F}_N(x^2)=\frac{1}{2^{2D}}\prod_{k=1}^{D}(x^2_k-x^2) \ee
It follows then that the factorized form of the eigenvalues is for even $N$:
\be \label{fact} D(u)=\frac{(D+1)}{2}2^{-2D}\prod_{k=1}^{D}(x_k+\epsilon_k x)(x_k+\mu_k x)  \ee
whereas for odd $N$:
\be D(u)=2^{-2D}\prod_{k=1}^D(x_k+\epsilon_k x)(x_k+\mu_k x)  \ee
being
\be  \mu_k^2=\epsilon_k^2=1  \ee
such solutions, however, are too many and one needs to impose some selection rules to pick the correct $(1,s)$ conformal boundary conditions.\\
The different sectors are chosen by applying selection rules to the combinatorics of zeroes.\\
A typical pattern  of zeroes for the eigenvalues for $N=12$ is (in the complex $u$ plane):
\psset{unit=.9cm}
\setlength{\unitlength}{.9cm}
\be
\begin{pspicture}(-.25,-.25)(14,12)
\psframe[linecolor=yellow,linewidth=0pt,fillstyle=solid,fillcolor=yellow](1,1)(13,11)
\psline[linecolor=black,linewidth=.5pt,arrowsize=6pt]{->}(4,0)(4,12)
\psline[linecolor=black,linewidth=.5pt,arrowsize=6pt]{->}(0,6)(14,6)
\psline[linecolor=red,linewidth=1pt,linestyle=dashed,dash=.25 .25](1,1)(1,11)
\psline[linecolor=red,linewidth=1pt,linestyle=dashed,dash=.25 .25](7,1)(7,11)
\psline[linecolor=red,linewidth=1pt,linestyle=dashed,dash=.25 .25](13,1)(13,11)
\psline[linecolor=black,linewidth=.5pt](1,5.9)(1,6.1)
\psline[linecolor=black,linewidth=.5pt](7,5.9)(7,6.1)
\psline[linecolor=black,linewidth=.5pt](10,5.9)(10,6.1)
\psline[linecolor=black,linewidth=.5pt](13,5.9)(13,6.1)
\rput(.5,5.6){\small $-\frac{\pi}{4}$}
\rput(6.7,5.6){\small $\frac{\pi}{4}$}
\rput(10,5.6){\small $\frac{\pi}{2}$}
\rput(12.6,5.6){\small $\frac{3\pi}{4}$}
\psline[linecolor=black,linewidth=.5pt](3.9,6.6)(4.1,6.6)
\psline[linecolor=black,linewidth=.5pt](3.9,7.2)(4.1,7.2)
\psline[linecolor=black,linewidth=.5pt](3.9,8.0)(4.1,8.0)
\psline[linecolor=black,linewidth=.5pt](3.9,9.0)(4.1,9.0)
\psline[linecolor=black,linewidth=.5pt](3.9,10.6)(4.1,10.6)
\psline[linecolor=black,linewidth=.5pt](3.9,5.4)(4.1,5.4)
\psline[linecolor=black,linewidth=.5pt](3.9,4.8)(4.1,4.8)
\psline[linecolor=black,linewidth=.5pt](3.9,4.0)(4.1,4.0)
\psline[linecolor=black,linewidth=.5pt](3.9,3.0)(4.1,3.0)
\psline[linecolor=black,linewidth=.5pt](3.9,1.4)(4.1,1.4)
\rput(3.6,6.6){\small $v_5$}
\rput(3.6,7.2){\small $v_4$}
\rput(3.6,8.0){\small $v_3$}
\rput(3.6,9.0){\small $v_2$}
\rput(3.6,10.6){\small $v_1$}
\rput(3.5,5.4){\small $-v_5$}
\rput(3.5,4.8){\small $-v_4$}
\rput(3.5,4.0){\small $-v_3$}
\rput(3.5,3.0){\small $-v_2$}
\rput(3.5,1.4){\small $-v_1$}
\psarc[linecolor=black,linewidth=.5pt,fillstyle=solid,fillcolor=black](1,6.6){.1}{0}{360}
\psarc[linecolor=gray,linewidth=0pt,fillstyle=solid,fillcolor=gray](1,7.2){.1}{0}{360}
\psarc[linecolor=black,linewidth=.5pt,fillstyle=solid,fillcolor=white](1,8.0){.1}{0}{360}
\psarc[linecolor=black,linewidth=.5pt,fillstyle=solid,fillcolor=black](1,9.0){.1}{0}{360}
\psarc[linecolor=gray,linewidth=0pt,fillstyle=solid,fillcolor=gray](1,10.6){.1}{0}{360}
\psarc[linecolor=black,linewidth=.5pt,fillstyle=solid,fillcolor=white](7,6.6){.1}{0}{360}
\psarc[linecolor=gray,linewidth=0pt,fillstyle=solid,fillcolor=gray](7,7.2){.1}{0}{360}
\psarc[linecolor=black,linewidth=.5pt,fillstyle=solid,fillcolor=black](7,8.0){.1}{0}{360}
\psarc[linecolor=black,linewidth=.5pt,fillstyle=solid,fillcolor=white](7,9.0){.1}{0}{360}
\psarc[linecolor=gray,linewidth=0pt,fillstyle=solid,fillcolor=gray](7,10.6){.1}{0}{360}
\psarc[linecolor=black,linewidth=.5pt,fillstyle=solid,fillcolor=black](13,6.6){.1}{0}{360}
\psarc[linecolor=gray,linewidth=0pt,fillstyle=solid,fillcolor=gray](13,7.2){.1}{0}{360}
\psarc[linecolor=black,linewidth=.5pt,fillstyle=solid,fillcolor=white](13,8.0){.1}{0}{360}
\psarc[linecolor=black,linewidth=.5pt,fillstyle=solid,fillcolor=black](13,9.0){.1}{0}{360}
\psarc[linecolor=gray,linewidth=0pt,fillstyle=solid,fillcolor=gray](13,10.6){.1}{0}{360}
\psarc[linecolor=black,linewidth=.5pt,fillstyle=solid,fillcolor=black](1,5.4){.1}{0}{360}
\psarc[linecolor=gray,linewidth=0pt,fillstyle=solid,fillcolor=gray](1,4.8){.1}{0}{360}
\psarc[linecolor=black,linewidth=.5pt,fillstyle=solid,fillcolor=white](1,4.0){.1}{0}{360}
\psarc[linecolor=black,linewidth=.5pt,fillstyle=solid,fillcolor=black](1,3.0){.1}{0}{360}
\psarc[linecolor=gray,linewidth=0pt,fillstyle=solid,fillcolor=gray](1,1.4){.1}{0}{360}
\psarc[linecolor=black,linewidth=.5pt,fillstyle=solid,fillcolor=white](7,5.4){.1}{0}{360}
\psarc[linecolor=gray,linewidth=0pt,fillstyle=solid,fillcolor=gray](7,4.8){.1}{0}{360}
\psarc[linecolor=black,linewidth=.5pt,fillstyle=solid,fillcolor=black](7,4.0){.1}{0}{360}
\psarc[linecolor=black,linewidth=.5pt,fillstyle=solid,fillcolor=white](7,3.0){.1}{0}{360}
\psarc[linecolor=gray,linewidth=0pt,fillstyle=solid,fillcolor=gray](7,1.4){.1}{0}{360}
\psarc[linecolor=black,linewidth=.5pt,fillstyle=solid,fillcolor=black](13,5.4){.1}{0}{360}
\psarc[linecolor=gray,linewidth=0pt,fillstyle=solid,fillcolor=gray](13,4.8){.1}{0}{360}
\psarc[linecolor=black,linewidth=.5pt,fillstyle=solid,fillcolor=white](13,4.0){.1}{0}{360}
\psarc[linecolor=black,linewidth=.5pt,fillstyle=solid,fillcolor=black](13,3.0){.1}{0}{360}
\psarc[linecolor=gray,linewidth=0pt,fillstyle=solid,fillcolor=gray](13,1.4){.1}{0}{360}
\end{pspicture}
\label{uplane}
\ee
A single zero is indicated by a grey dot while a double zero is indicated by
a black dot.
\subsection{Selection Rules}
A two column configuration is a couple $(\vec{l}|\vec{r})$ of vectors both of length $M$ with integral entries  arranged in decreasing order.\\
A two column configuration is called \emph{admissible} if, calling $m$ the length of $\vec{l}$ one has:
\be l_k\leq r_k ,\ k=1,\ldots,m \ee
It follows then that to each zero pattern of the eigenvalues it is possible to associate only a single two-column configuration can be described as described in figure \ref{onetwo}, where one is describing the state $(3|4,3,1)$ .\\
\psset{unit=.7cm}
\setlength{\unitlength}{.7cm}
\be
\begin{pspicture}(-.25,-.25)(2,5)
\psframe[linewidth=0pt,fillstyle=solid,fillcolor=yellow](0,0)(2,5)
\psarc[linecolor=black,linewidth=.5pt,fillstyle=solid,fillcolor=white](1,4.5){.1}{0}{360}
\psarc[linecolor=gray,linewidth=0pt,fillstyle=solid,fillcolor=gray](1,3.5){.1}{0}{360}
\psarc[linecolor=black,linewidth=.5pt,fillstyle=solid,fillcolor=black](1,2.5){.1}{0}{360}
\psarc[linecolor=black,linewidth=.5pt,fillstyle=solid,fillcolor=white](1,1.5){.1}{0}{360}
\psarc[linecolor=gray,linewidth=0pt,fillstyle=solid,fillcolor=gray](1,0.5){.1}{0}{360}
\end{pspicture}
\hspace{.6cm}\ \ \longleftrightarrow \hspace{.6cm}
\begin{pspicture}(-.25,-.25)(2,5)
\psframe[linewidth=0pt,fillstyle=solid,fillcolor=yellow](0,0)(2,5)
\psarc[linecolor=black,linewidth=.5pt,fillstyle=solid,fillcolor=white](0.5,4.5){.1}{0}{360}
\psarc[linecolor=black,linewidth=.5pt,fillstyle=solid,fillcolor=white](0.5,3.5){.1}{0}{360}
\psarc[linecolor=gray,linewidth=0pt,fillstyle=solid,fillcolor=gray](0.5,2.5){.1}{0}{360}
\psarc[linecolor=black,linewidth=.5pt,fillstyle=solid,fillcolor=white](0.5,1.5){.1}{0}{360}
\psarc[linecolor=black,linewidth=.5pt,fillstyle=solid,fillcolor=white](0.5,0.5){.1}{0}{360}
\psarc[linecolor=black,linewidth=.5pt,fillstyle=solid,fillcolor=white](1.5,4.5){.1}{0}{360}
\psarc[linecolor=gray,linewidth=0pt,fillstyle=solid,fillcolor=gray](1.5,3.5){.1}{0}{360}
\psarc[linecolor=gray,linewidth=0pt,fillstyle=solid,fillcolor=gray](1.5,2.5){.1}{0}{360}
\psarc[linecolor=black,linewidth=.5pt,fillstyle=solid,fillcolor=white](1.5,1.5){.1}{0}{360}
\psarc[linecolor=gray,linewidth=0pt,fillstyle=solid,fillcolor=gray](1.5,0.5){.1}{0}{360}
\end{pspicture}
\label{onetwo}
\ee
The label $k$ in $l_k, r_k$ is \emph{the same} $k$ as in \ref{fact} (we will understand this better from the IOM) and one has:
\be  \epsilon_n=-1, \ {\rm if}\ n\in \{l_1,\ldots,l_M\}, \ \epsilon_n=1 \ {\rm otherwise}  \ee
\be  \mu_n=-1, \ {\rm if}\ n\in \{r_1,\ldots,r_M\}, \ \mu_n=1 \ {\rm otherwise}  \ee
We recall from \cite{solvpol} that the set $A_{m,n}^{M}$ is the set of all \emph{admissible} two column diagrams of height $M$ with $m$ occupied sites on the left and  $n$ occupied sites on the right.\\ 
To each two column diagram ${\cal D}$ is associated a weight:
\be w({\cal D})=\sum_i l_i+\sum_j r_j  \ee
one then defines:
\be \nar{M}{m}{n}=\sum_{{\cal D}\in A_{m,n}^M}q^{w({\cal D})}   \ee
\be \nar{M}{m}{n}=0 , \quad { \rm if} \ A_{m,n}^M=\emptyset   \ee
one then has the following Fermionic formuals for the finitized characters \cite{solvpol}.\\
For odd $s$ one has:
\be \chi_{1,s}^{(N)}(q)=q^{\frac{1}{12}}\sum_{m=0}^{\frac{N-s+1}{2}}\Big(\nar{\frac{N}{2}}{m}{m+\frac{s-3}{2}}+\nar{\frac{N-2}{2}}{m}{m+\frac{s-1}{2}}\Big)   \ee
For even $s$, one has:
\be   \chi_{1,s}^{(N)}(q)=q^{-\frac{1}{24}-\frac{s-2}{4}}\sum_{m=0}^{\frac{N-s+1}{2}}\nar{\frac{N-1}{2}}{m}{m+\frac{s-2}{2}} q^{-m}    \ee
Clearly $\nar{M}{m}{n}$ is the character associated to the set $A_{m,n}^M$ with respect to the weight introduced above.\\ 
From these expressions one can read off at first sight which two column diagrams are allowed to contribute to a given sector.

\section{Transfer Matrix Expansion on the Temperley Lieb Algebra}
We are now interested in providing an expansion for the transfer matrix over a basis of suitable $N-$tangles defined in the TL algebra, indeed in \cite{nigropol} it was widely discussed how an expansion in terms of $x$ can be explicitly obtained which looks like:
\be \label{susu}\D(u)=\uno+\sum_{n=1}^{2D} {\bf\hat D}_n \sin^n(2u) \ee
This decomposition is reached  by first starting from the more natural  expansion, which is readily obtained from the expansion of the elementary faces in terms of connecions:
\be \D(u)=\frac{1}{2}\sum_{k=1}^{2N-1}\cos^{2N-k-1}(u)\sin^{k-1}(u)\D_k     \ee
The tangles $\hat\D_n$ are related to the $\D_n$ by:
\be\label{wazza} {\bf\hat D}_n= \frac{1}{2^{n+1}}\Big\{{\bf D}_{n+1}+\sum_{j=1}^{\lfloor\frac{n}{2}\rfloor}\frac{(-1)^j(N-n-1+2j)}{j}\binom{N-n-2+j}{j-1}{\bf D}_{n+1-2j}\Big\}  \ee
and
\be\begin{split} {\bf D}_{n}&=\sum_{\mu=1}^{N-1}\sum_{d,\alpha,\beta\geq 0}\delta_{2\mu+d+\alpha+\beta,n+1}\\
&\cdot\Big\{{\bf e}_{1}\ldots{\bf e}_{\mu+d-1}\Big(\sum_{\mu+d<i_1<\ldots<i_\alpha\leq N-1}{\bf e}_{1}\ldots {\bf e}_{i_\alpha}\Big)\Big(\sum_{\mu<j_1<\ldots<j_\beta\leq N-1}{\bf e}_{j_\beta}\ldots {\bf e}_{j_1}\Big)\Big\}+\\
&+\Big(\sum_{\mu<i_1<\ldots<i_\alpha\leq N-1}{\bf e}_{1}\ldots {\bf e}_{i_\alpha}\Big)\Big(\sum_{\mu+d<j_1<\ldots<j_\beta\leq N-1}{\bf e}_{j_\beta}\ldots {\bf e}_{j_1}\Big){\bf e}_{1}\ldots{\bf e}_{\mu+d-1}\Big\} \end{split}\ee
The expressions for the first $\D_n$ read:
\be\begin{split}    {\bf D}_1&= {\bf B}_0     \\
        {\bf D}_2&= 4\sum_{j=1}^{N-1}{\bf e}_j \\
        {\bf D}_3&= {\bf B}_0 +4\sum_{1\leq i<j\leq N-1}\{{\bf e}_i,{\bf e}_j\}   \\
        {\bf D}_4&= 4 {\bf B}_1+8\sum_{j=2}^{N-2}{\bf e}_j+4\sum_{1\leq i<j<k\leq N-1}\{{\bf e}_i,\{{\bf e}_j,{\bf e}_k\}\}  \\
        {\bf D}_5&= {\bf B}_0+4\sum_{2\leq i\leq N-2}\{{\bf e}_i,{\bf B}_1\}+8\sum_{2\leq i<j\leq N-2}\{{\bf e}_i,{\bf e}_j\}+4\sum_{1\leq i<_2j\leq N-1}\{{\bf e}_i,{\bf e}_j\}+\\
        &+4\sum_{1\leq i<j<k<l\leq N-1}\{{\bf e}_i,\{ {\bf e}_j,\{ {\bf e}_k, {\bf e}_l \}\}\}   \\
        \ldots &       \end{split}\ee
 where
 \be a<_n b \Longleftrightarrow b-a\geq n\ee
These tangles will be the building blocks for the expansion of the ${\bf Q}$ operator of Critical Dense Polymers.\\

\section{Some Common Lore on T-Q Relations in CFT and Local and Nonlocal Conservation Laws}
This section has basicly the goal to collect some known things from the works \cite{blz} with the aim to make this paper more self contained and have a more clear idea of which objects we will be actually rebuilding in the Temperley Lieb algebra and comparing to their continuum relatives.\\
It is well known that in CFT some operators ${\bf T}_j(\lambda)$ with $j=0,1/2,1,3/2,\ldots$ can be built as traces of monodromy operators over auxiliary vector spaces $\mathbb{C}^{2j+1}$, which form commuting families in the spectral parameter $\lambda$. These monodromy operators are easily built in the Feigin-Fuchs Free Field realization of the Virasoro algebra \cite{blz} which involves a single free bosonic field. These operators act invariantly on Verma Modules of central charge and conformal weight:
\be c=13-6(\beta^2+\beta^{-2})    \ee
\be \Delta= \big(\frac{p}{\beta}\big)^2+\frac{c-1}{24}    \ee
where $p$ is the eigenvalue of the momentum operator $P$ which appears in the zero mode Heisenberg algebra of the bosonic free field.\\
They satisfy fusion relations:
\be {\bf T}_{\frac{1}{2}}(\lambda){\bf T}_j(q^{j+\frac{1}{2}}\lambda)={\bf T}_{j-\frac{1}{2}}(q^{j+1}\lambda)+{\bf T}_{j+\frac{1}{2}}(q^{j}\lambda)    \ee 
being
\be  q=e^{i\pi \beta^2}   \ee
in particular we are more interested in this paper in the operator $T_{\frac{1}{2}}(\lambda)=T(\lambda)$ which we can consider as the quantum version of the lattice transfer matrix.\\
The quantum transfer matrix is related to the infinite dimensional abelian algebra of local conserved charges ${\bf I}_{2n-1}$, which make the theory integrable, via the following asymptotic expansion:
\be  \log {\bf T}(\lambda)\sim m \lambda^{\frac{1}{1-\beta^2}}\uno-\sum_{n=1}^\infty C_n\lambda^{\frac{1-2n}{1-\beta^2}}{\bf I}_{2n-1}     \ee
for some suitable constants $C_n$.\\
There is also another family of commuting operators ${\bf Q}(\lambda)$ which is defined by the well known $T-Q$ relations:
\be {\bf T}(\lambda){\bf Q}(\lambda)={\bf Q}(q \lambda)+{\bf Q}(q^{-1}\lambda)  \ee
these operators also act invariantly on the assigned Verma Modules, and can be used to define nonlocal conserved charges ${\bf H}_n$ in conformal field theory (the nonlocality being due to appearence of  integrated products of screening operators at different points in their bosonic representation ):
\be  \log{\bf Q}\sim\frac{2 P}{\beta^2}\log \lambda-\sum_{n=1}^\infty \Big(\frac{\Gamma(1-\beta^2)}{\beta^2}\Big)^{2n}\lambda^{2n}{\bf H}_n       \ee
also, it is possible to define dual nonlocal conserved charges $\tilde{\bf H}_n$ whose definition is again found in \cite{blz}. We remand to those works for this definition.\\
All the information on conserved charges in conformal field theory can be encoded into a function which is a suitable integral tranform of the ${\bf Q}$ operator:
\be {\bf \Psi}(\nu)=K(\nu,\beta^2)\int_{-\infty}^0\frac{d\lambda^2}{\lambda^2}(-\lambda^2)^{-i\nu(1+\xi)/2}\log{\bf Q}(\lambda)   \ee
for some function $K(\nu,\beta^2)$ which can be read off in the original work, and after all the really important information about CFT is contained inside the integral term.\\
Throughout this work we will be especially interested in the situation where $P=0$, basicly for the sake of being clear and simple.\\
This special function ${\bf \Psi}$ is found, at specific values on the imaginary axis, to take the same values as the local and nonlocal conserved charges of conformal field theory. For example:
\be {\bf  \Psi}((2n-1)i)=\beta^{2n}{\bf I}_{2n-1}   \ee 
we omit the other relations and refer the reader to the original works, asking forgiveness for not being totally self contained.\\

\section{Baxter's Q}
We want to consider the problem of explicitly characterizing the ${\bf Q}$ operator of critical dense polymers in terms of N-tangles defined in the temperley lieb algebra with vanishing loop fugacity.\\
We recognize that the variables $\lambda, q$ of \cite{blz} (BLZ) take the values $q=i$ and $\lambda^2=\sin(2u)=x$, and notice that the transfer matrix ${\bf T}$ and the ${\bf Q}$ operator must be considered as analytic functions of $x$, it follows then that the T-Q relation for the ${\bf T},{\bf Q}$ in critical dense polymers must be set in the following form:
\be {\bf T}(x){\bf Q}(x)={\bf Q}(q^2 x)+{\bf Q}(q^{-2}x)=2{\bf Q}(-x)   \ee
this equation, however can admit a nontrivial solution for $\bf Q$ only if the tranfer matrix satisfies a suitably normalized inversion identity:
\be {\bf T}(x){\bf T}(-x)=4\uno  \ee
The normalized ${\bf T}$ can be written for odd $N$ as:
\be  {\bf T}(x)=\frac{2}{{\cal F}(x)}{\bf D}(x)  \ee
whereas for even $N$:
\be {\bf T}(x)=\frac{4}{{\cal F}(x)}{\bf D}(x)  \ee
where:
\be {\cal F}(x)= 1+\sum_{n=1}^{2D} F_{2n}x^{2n} \ee
and $N=2D+2$ for even $N$, whereas $N=2D+1$ for odd $N$, and also $x=\sin(2u)$.\\
One then decides to introduce the following expansion:
\be  {\bf T}(x)=2+\sum_{n=1}^{2D} {\bf T}_n x^n   \ee
Indeed one can be even more precise and relate the tangles ${\bf T}_n$ to the tangles $\hat{\bf D}_n$ as:
\be {\bf T}_1=2\hat {\bf D}_1   \ee 
\be {\bf T}_2=2\hat {\bf D}_2-F_2\uno   \ee
\be {\bf T}_3=2\hat {\bf D}_3 -2F_2\hat {\bf D}_1 \ee
\be {\bf T}_4= 2\hat {\bf D}_4-2F_2\hat{\bf D}_2+2(F_2^2-F_4)\uno       \ee
\be \ldots \ee
notice also that:
\be F_{2k}=\frac{\alpha_D}{ 2^{2D}}\sum_{1\leq k_1<\ldots,<k_D\leq D}x_{2i_1}\ldots x_{2i_n}   \ee
where $\alpha_D=1$ for odd $N=2D+1$ and $\alpha_D=D+1$ for even $N=2D+2$.\\
Furthermore the eigenvalues of the transfer matrix can be expressed, in view of the selection rules as:
\be  T(x)= 2\prod_{k=1}^D \frac{x_k-x}{x_k+x} \prod_{k\in {\cal D}}\frac{x_k+x}{x_k-x}   \ee
Such a factorized form is simply reproduced by a suitable factorized form for the eigenvalues of the operator ${\bf Q}$.\\
The T-Q relation can be solved recursively for the coefficients ${\bf Q}_n$ of the expansion:
\be {\bf Q}(x)=1+\sum_{n=1}^\infty {\bf Q}_{n} x^n  \ee
yielding the recursion:
\be \label{rec} 2((-1)^l-1){\bf Q}_l=\sum_{k=0}^{l-1}{\bf T}_{l-k}{\bf Q}_k  \ee
which tells us explicitly that, 
\be {\bf Q_1 }=-\frac{1}{4}{\bf T}_1   \ee
\be {\bf T}_2=\frac{1}{4}{\bf T}_1^2  \ee
\be {\bf Q}_3=-\frac{1}{4}{\bf T}_3+\frac{1}{16}{\bf T}_1^3-\frac{1}{4}{\bf T}_1{\bf Q}_2  \ee
\be {\bf T}_4=\frac{1}{2}{\bf T}_1{\bf T}_3-\frac{1}{64}{\bf T}_1^4-\frac{1}{4}{\bf T}_1^2{\bf Q}_2  \ee
\be {\bf Q}_5=-\frac{1}{4}{\bf T}_5+\frac{3}{64}{\bf T}_1^2{\bf T}_3-\frac{1}{512}{\bf T}_1^5+\frac{1}{64}{\bf T}_1^3{\bf Q}_2-\frac{1}{4}{\bf T}_3 {\bf Q}_2-\frac{1}{4}{\bf T}_1{\bf Q}_4  \ee
\be \ldots   \ee
we notice that these equations are unable to fix the ${\bf Q}_{2n}$ in a simple linear way, this has to do with the fact that we are free to rescale ${\bf Q}$ by an arbitrary factor which is invariant under $x\to-x$. For this reason we choose to set:
\be  {\bf Q}_{2n}= Q_{2n}\uno  \ee
We now switch our interest on the consequences of knowing the exact values of the roots for the eigenvalues of the transfer matrix ${\bf T}$. As a consequence of the simple T-Q relations at $c=-2$, one has that by assuming a motivated ansatz for the eigenvalues of Q
\be Q(x)=\prod_{k=1}^D \Big(1+\frac{x}{x_k}\Big)\prod_{k\in {\cal D}}\Big(1-\frac{x}{x_k}\Big) \label{fac}  \ee
we notice that the $Q$ satisfy the following equation:
\be  \frac{Q(x)}{Q(-x)}=\prod_{k=1}^D \frac{x_k-x}{x_k+x} \prod_{k\in {\cal D}}\frac{x_k+x}{x_k-x} \ee 
it follows from the $T-Q$ relation that the $x_k$ \emph{are} the zeroes of $T$, thus
\be  x_k=\frac{1 }{\sin t_k} \ee
It follows that our ansatz for $Q$ provides a suitable solution for the $T-Q$ relations, and that the zeroes of $Q$ are known exactly.\\ 
Somehow the simplicity of the Bethe equations for Critical Dense Polymers, is related to the fact that the Symplectic Fermion description of the model somehow is exact on the lattice as well (at least the combinatorics in characters is manifestly the same), on the subject see for example \cite{salLCFT}\cite{solvpol}\cite{nigropol}.

\section{Integrals of Motion and Lattice ${\bf \Psi}$ Function}
We consider now, inspiring ourselves once more to \cite{blz} the following matrix valued function of the complex variable $\nu$:
\be {\bf \Psi}(\nu)=\frac{\nu\sin(\pi\nu)}{\pi}N^{-i\nu}\int_{-\infty}^0\frac{dx}{x}(-x)^{-i\nu}\log{\bf Q}(x)   \ee
where if we consider the corresponding eigenvalue problem, the above integral is convergent for $0<{\rm Im}(\nu)<1$, outside this domain we shall consider the function ${\bf \Psi}(\nu)$ to be defined by analytic prolongation from the region where the integral is convergent.\\
We recall that the ${\bf Q}_{2n-1}$ are expressed in terms of the tangles ${\bf T}_n$ by means of the recursion \ref{rec}.\\ 
It follows from the factorization property \ref{fac} for the eigenvalues of  ${\bf Q}(x)$ that the eigenvalues of the matrix valued function ${\bf \Psi}(\nu)$ are given by:
\be \Psi(\nu)=N^{-i\nu}(\sum_{k\in{\cal D}}(\sin t_k)^{-i\nu} +\sum_{k=1}^D(-\sin t_k)^{-i\nu} ) \ee
it is now immediate to compute the limit of the eigenvalues as $N\to\infty$:
\be \Psi(\nu)\sim (\pi)^{-i\nu}(\sum_{k\in{\cal D}}((k+a))^{-i\nu}+\sum_{k=1}^D((k+a))^{-i\nu}+O(N^{-2}))  \ee
and since
\be \sum_{k=1}^D(-(k+a))^{-i\nu}\sim (-1)^{-i\nu}\zeta(i\nu,a)+O(1/N) \ee
clearly they are proportional to the eigenvalues of the corresponding ${\bf \Psi}$ operator in the continuum which was introduced in \cite{blz} BLZ, and $a=\frac{1}{2}$ for odd $N$ whereas $a=0$ for even $N$.\\
Actually one has:
\be \Psi(\nu)\sim\frac{(\pi)^{-i\nu}}{2^{i\nu-1}(i\nu-1)}\Psi_{BLZ}(\nu)+O(N^{-1})\ee
where 
\be  \Psi_{BLZ}(\nu)=2^{i\nu-1}(i\nu-1)(\zeta(i\nu,a)+\sum_{k\in {\cal D}}(a+k)^{-i\nu})   \ee
actually it is well known that ${\bf \Psi}_{BLZ}(\nu)$ generates all the local, non local and dual non local IOM, therefore one can consider the above definition on the lattice to give rise to local and nonlocal charges on the lattice as well, by means of the definitions:
\be {\bf \Psi}((2n-1)i)={\bf I}_{2n-1}   \ee
Where one has that:
\be \log{\bf T}(x)=\log 2 +\sum_{n=1}^\infty \frac{Y_n(\frac{1!}{2}{\bf T}_1,\ldots,\frac{n!}{2}{\bf T}_n)}{n!}x^n  \ee
where we introduced the inverse Bell Polynomials, which are defined by:
\be \log(1+\sum_{k=1}^\infty \frac{C_k}{k!}x^k)=\sum_{k=1}^\infty \frac{Y_k(C_1\ldots,C_k)}{k!}x^k        \ee
defined by recurrence as:
\be Y_{n+1}(C_1,\ldots,C_{n+1})=C_{n+1}-\sum_{k=1}^{n}\binom{n}{k-1}C_{n-k+1}Y_{k}(C_1,\ldots,C_k)  \ , \qquad Y_0=1\ee
clearly one has for some proportionality constant:
\be  {\bf I}_{2n-1}\sim Y_n(\frac{1!}{2}{\bf T}_1,\ldots,\frac{n!}{2}{\bf T}_n)  \ee
Ultimately the local lattice IOM ${\bf I}_{2n-1}$ are closely related to the tangles ${\bf K}_{2n-1}$ of \cite{nigropol}, indeed we now that for fixed size only a finite number of nondegenerate ${\bf K}_{2n-1}$ is allowed to exist, for this reason the first ${\bf I}_{2n-1}$ should be proportional to the ${\bf K}_{2n-1}$, for high enough $n$ however the ${\bf I}_{2n-1}$ take degenerate  expressions which follow from ${\bf D}(x)$  being a polynomial instead of a full power series.\\
We now turn our attention to the non local conserved charges, they are defined through
\be  {\bf \Psi}(-in)= n N^{-n}{\bf H}_n   \ee
the eigenvalues of the ${\bf H}_n$ are found to be:
\be H_n=\frac{1}{n}(\sum_{k\in{\cal D}}x_k^{-n}+\sum_{k=1}^D(-x_k)^{-n})   \ee
in the limit $N\to \infty$ they are proportional to the eigenvalues of the nonlocal IOM ${H}_n$ of BLZ.\\
The nonlocal IOM ${\bf H}_n$ are also understood to arise from the following expansion for the ${\bf Q}$ operator:
\be \log{\bf Q}(x)= -\sum_{n=1}^\infty {\bf H}_n x^n   \ee
where 
\be {\bf H}_n=\frac{Y(1!{\bf Q}_1,\ldots,n!{\bf Q}_n)}{n!}  \ee
notice that :
\be \log Q(x)=\sum_{j\in{\cal D}}\log\Big(1-\frac{x}{x_j}\Big)+\sum_{j=1}^D\log\Big(1+\frac{x}{x_j}\Big)=-\sum_{k=1}^\infty x^k \frac{1}{k}(\sum_{j\in {\cal D}}\frac{1}{x_j^k}+\sum_{j=1}^D\frac{1}{(-x_j)^k}\Big)  \ee
and thus we recover again:
\be H_n=\frac{1}{n}(\sum_{j\in{\cal D}}\frac{1}{x_j^n}+\sum_{j=1}^D\frac{1}{(-x_j)^n})   \ee 
We have thus introduced an analytic prolongation of the local and non local IOM to arbitrary complex values of their index $n$ (ultimately related to their spin), and shown how it is described in a simple continuum limit by the $\Psi$ function of BLZ. This useful tool encodes all the information on conserved quantities in Critical Dense Polymers, and by means of an inverse integral transform it can be related both to $T$ and $Q$.\\
\section{Continuum Limit}
It is well known that a suitable continuum limit can be obtained for Critical Dense Polymers by letting the spectral parameter go into the braid limit $u\to i\infty$ and simultaneously letting the system size $N$ go to infinity, in such a way that the following combination is kept constant:
\be \hat x= \frac{\pi \sin(2u)}{N}  \ee
in this limit the eigenvalues of the ${\bf T},{\bf Q}$ operators become:
\be  T(\hat x)=2\prod_{k=1}^\infty \frac{k+a-\hat x}{k+a+\hat x} \prod_{k\in {\cal D}}\frac{k+a+\hat x}{k+a-\hat x} \ee
\be Q(\hat x)=\prod_{k=1}^\infty \Big(1+\frac{\hat x}{k+a}\Big)\prod_{k\in {\cal D}}\Big(1-\frac{\hat x}{k+a}\Big)   \ee
We notice that $Q(\hat x)$ is formally porportional to the spectral determinant of the harmonic oscillator, which however when naively defined does not give rise to convergent products, so that one needs to introduce Weierstrass factors in the product definition to make sense out of them. We start with:
\be  \prod_{k=1}^\infty \Big(1+\frac{\hat x}{k+a}\Big)e^{-\frac{{\cal C}\hat x}{\gamma n}}= e^{-{\cal C}\hat x} \frac{\Gamma(1+a)}{\Gamma(1+a-\hat x)}   \ee
where $\gamma $ is Euler's constant, and ${\cal C}$ is a renormalization constant. Similarly one has:
\be \prod_{k=1}^\infty \Big(\frac{k+a+\hat x}{k+a-\hat x}\Big)e^{2\frac{{\cal C}\hat x}{\gamma n}}= e^{2{\cal C}\hat x}\frac{\Gamma(1+a+\hat x)}{\Gamma(1+a-\hat x)} \ee
so the continuum expressions for the eigenvalues $T,Q$ are:
\be  T(\hat x)=2 e^{2{\cal C}\hat x} \frac{\Gamma(1+a
-\hat x)}{\Gamma(1+a+\hat x)} \prod_{k\in {\cal D}}\frac{k+a+\hat x}{k+a-\hat x} \ee
\be Q(\hat x)=e^{-{\cal C}\hat x}\frac{\Gamma(1+a)}{\Gamma(1+a-\hat x)}\prod_{k\in {\cal D}}\Big(1-\frac{\hat x}{k+a}\Big)   \ee
which are results well known in the continuum since the times of \cite{blz}.\\
Similarly analisys of the eigenvalues allos to estabilish that the continuum limit of the operator ${\bf \Psi}(\nu)$ is proportional to the operator ${\bf \Psi}_{BLZ}(\nu)$ defined in Conformal Field Theory:
\be {\bf \Psi}(\nu)=\frac{(\pi)^{-i\nu}}{2^{i\nu-1}(i\nu-1)}{\bf \Psi}_{BLZ}(\nu)    \ee
From this it follows that also all the local and non local charges must go in the continuum limit to expressions proportional to the operators of BLZ.\\

\section{Discussion}
In this paper we have explicitly built the ${\bf Q}$ operator of Critical Dense Polymers from the Temperley Lieb Algebra, and obtained expressions for the local and non local involutive charges on the lattice, these results \emph{are new} (at least if we neglect the author's own work \cite{nigropol} on the local charges on the lattice), even if the methods employed date back to \cite{blz}, and therefore this paper should not be considered only just as a pedagogical exercise, even if it does not look particularly complex when compared to the existing literature on the subject (the author hopes this will at least increase the number of his readers!!). These lattice definitions are found to behave properly in the continuum limit and to reproduce the corresponding quantities of Conformal Field Theory.\\
But this is not all the story, we must confess however that we have been lazy, because we could have computed the full $1/N$ expansion for the lattice involutive charges, again in the spirit of \cite{nigropol}, also for the nonlocal charges to find out that, once again, each term in this expansion can be uniquely expressed in terms of conformally invariant quantities. This aspect, to the author's opinion quite fundamental (and holding also for the ising model in the RSOS formulation), may have passed unnoticed in \cite{nigropol}, being only one of the issues addressed there among many. This feature in turn is suspected to be a completely generaly feature of integrable models close to their critical points which admit a transfer matrix description. Of course a completely general proof of this conjecture is not yet existing. It would be nice if some interested reader tried to tackle this question himself for other values of the central charge. Of course due to lack of exact knowledge of the eigenvalues of the involutive charges the most natural method to use for this investigation would be a numerical one.\\
All the results in this paper should be easily generalizable to the Ising Model, by using a different decomposition for the lattice operators which is given by the standard basis for the Clifford algebra of $\gamma$ matrices. In this case however the general expression for the eigenvalues of the ${\bf \Psi}$ operator of conformal field theory are not known in general, although the author currently has conjectural exact expressions for these objects in the case of the Ising model. A future project is to test these expressions against for example \cite{spdet}, the check of course has to be numerical and basicly is reduced to computing spectral zeta functions of the Shroedinger operators defined there.\\
Also, in the case of Ising it is possible to obtain the massive flow of the nonlocal charges towards the IR $c=0$ fixed point under a bulk thermal perturbation.\\

\end{document}